\documentclass[letterpaper]{article} % DO NOT CHANGE THIS
\usepackage{aaai2026}  % DO NOT CHANGE THIS
\usepackage{times}  % DO NOT CHANGE THIS
\usepackage{helvet}  % DO NOT CHANGE THIS
\usepackage{courier}  % DO NOT CHANGE THIS
\usepackage[hyphens]{url}  % DO NOT CHANGE THIS
\usepackage{graphicx} % DO NOT CHANGE THIS
\usepackage{natbib}  % DO NOT CHANGE THIS AND DO NOT ADD ANY OPTIONS TO IT
\usepackage{caption} % DO NOT CHANGE THIS AND DO NOT ADD ANY OPTIONS TO IT
\usepackage{algorithm}
\usepackage{algorithmic}
\usepackage{csquotes}
\MakeOuterQuote{"}
\usepackage{booktabs,tabularx,array} 

\usepackage{newfloat}
\usepackage{listings}
\DeclareCaptionStyle{ruled}{labelfont=normalfont,labelsep=colon,strut=off} % DO NOT CHANGE THIS
\floatstyle{ruled}
\newfloat{listing}{tb}{lst}{}
\floatname{listing}{Listing}
\usepackage{hyperref} %-- This package is specifically forbidden

\title{Inter-Agent Trust Models: A Comparative Study of Brief, Claim, Proof, Stake, Reputation and Constraint in Agentic Web Protocol Design—A2A, AP2, ERC-8004, and Beyond}

\author {
    Botao `Amber' Hu\textsuperscript{\rm 1},
    Helena Rong\textsuperscript{\rm 2}
}
\affiliations{
    \textsuperscript{\rm 1}University of Oxford, \textsuperscript{\rm 2} New York University Shanghai\\
    \textsuperscript{\rm 1} botao.hu@cs.ox.ac.uk, 
    \textsuperscript{\rm 2} hr2703@nyu.edu
}

\usepackage{bibentry}
\begin{document}

\maketitle

\begin{abstract}
As the “agentic web” takes shape—billions of AI agents (often LLM-powered) autonomously transacting and collaborating—trust shifts from human oversight to protocol design. In 2025, several inter‑agent protocols crystallized this shift, including Google’s Agent‑to‑Agent (A2A), Agent Payments Protocol (AP2), and Ethereum’s ERC‑8004 “Trustless Agents,” yet their underlying trust assumptions remain under‑examined. This paper presents a comparative study of trust models in inter‑agent protocol design: Brief (self‑ or third‑party verifiable claims), Claim (self-proclaimed capabilities and identity, e.g. AgentCard), Proof (cryptographic verification, including zero‑knowledge proofs and trusted execution environment attestations), Stake (bonded collateral with slashing and insurance), Reputation (crowd feedback and graph‑based trust signals), and Constraint (sandboxing and capability bounding). For each, we analyze assumptions, attack surfaces, and design trade‑offs, with particular emphasis on LLM‑specific fragilities—prompt injection, sycophancy/nudge‑susceptibility, hallucination, deception, and misalignment—that render purely reputational or claim‑only approaches brittle. Our findings indicate no single mechanism suffices. We argue for trustless‑by‑default architectures anchored in Proof and Stake to gate high‑impact actions, augmented by Brief for identity and discovery and Reputation overlays for flexibility and social signals. We comparatively evaluate A2A, AP2, ERC‑8004 and related historical variations in academic research under metrics spanning security, privacy, latency/cost, and social robustness (Sybil/collusion/whitewashing resistance). We conclude with hybrid trust model recommendations that mitigate reputation gaming and misinformed LLM behavior, and we distill actionable design guidelines for safer, interoperable, and scalable agent economies.
\end{abstract}


\section{Introduction}
The emergence of an "agentic web" of AI—potentially billions of autonomous agents interacting online—poses a fundamental challenge: how can these agents reliably trust one another without direct human supervision \cite{Yang2025Agenticb}? Traditional internet trust mechanisms (e.g., DNS names, TLS certificates) assume relatively static, human-operated services and an ownership-based trust model, which cannot meet the millisecond-by-millisecond dynamic coordination and verification needs of large-scale AI agentic ecosystems \cite{Raskar2025DNS}. In this new paradigm, trust is increasingly determined by the protocols that govern agent interactions, rather than by human judgment or centralized authorities. Ensuring robust trust among AI agents is critical because these agents will be entrusted with sensitive tasks—financial transactions, personal data handling, critical infrastructure control—where errors or abuse could have serious consequences \cite{Yang2025Agenticb}. Recent studies of AI safety underscore this urgency: even state-of-the-art large language model (LLM)-based agent frameworks exhibit fragilities and untrustworthiness such as prompt injection \cite{Liu2024Prompt}, sycophancy \cite{Sharma2025Understanding}, biases \cite{Rettenberger2025Assessing}, nudge-susceptibility \cite{Cherep2025LLM}, hallucination \cite{Xu2025Hallucination}, deception \cite{Hubinger2024Sleeper}, and misalignment \cite{Shen2023Large}, indicating unresolved trust issues in coordination, reliability, and oversight. Addressing inter-agent trust is thus pivotal for achieving safe autonomy at scale \cite{Yang2025Agenticb}.

In 2025, a number of open protocols were proposed to establish common standards for agent-to-agent interaction and trust. For example, Google’s Agent-to-Agent (A2A) communication protocol enables autonomous agents to discover each other and collaborate across organizational boundaries via standardized skill advertisements (AgentCards) and secure messaging \cite{a2a}. Similarly, the Agent Payments Protocol (AP2) was introduced by Google, PayPal and others to let AI agents perform financial transactions on behalf of users, with an emphasis on auditable, accountable payment flows underpinned by cryptographic user intent proofs \cite{ap2}. In the blockchain space, Ethereum’s proposed ERC‑8004 “Trustless Agents” \cite{erc8004} standard seeks to leverage on-chain registries for agent identity, reputation, and validation, allowing agents to be discovered and chosen “without pre-existing trust” . Each of these initiatives encapsulates design assumptions about how agents establish trust—be it through reputational feedback, credential verification, economic incentives, or sandboxed constraints. However, there has been little systematic analysis of the trust models underlying these protocols.

In this paper, we identify and compare six distinct trust models employed (implicitly or explicitly) in inter-agent protocol design: \textit{Brief}, \textit{Claim}, \textit{Proof}, \textit{Stake}, \textit{Reputation}, and \textit{Constraint}. Brief-based trust relies on endorsed claims or certificates (e.g. verifiable credentials) that an agent or third parties provide as evidence of identity or capability. Claim-based trust is the agent’s self-proclaimed identity and abilities (e.g. an AgentCard describing what the agent can do), without external verification. Proof-based trust requires cryptographic or formal proofs of behavior or state, such as digital signatures, zero-knowledge proofs of correct computation, or TEE remote attestations that vouch for an agent’s code integrity  . Stake-based trust involves economic skin-in-the-game: agents put down collateral that can be slashed for misbehavior, sometimes supplemented by insurance pools to cover damages . Reputation-based trust aggregates feedback or ratings from other agents/users into trust scores, or constructs trust graphs to inform decision-making. Finally, Constraint-based trust is achieved by sandboxing and bounding an agent’s actions and access, so that even a misaligned or malicious agent is technically limited in the harm it can do.

Overall, our study finds that no single trust mechanism is sufficient for the complexities of open multi-agent environments. Simple reputation systems or unsigned agent claims leave too many attack surfaces (Sybil attacks, collusion, lying agents), whereas purely cryptographic approaches can be costly or impractical for real-time agent orchestration. The optimal design appears to be hybrid: defaulting to minimal trust (zero-trust principles) for high-impact actions, but opportunistically layering trust signals (credentials, stake, reputation) as additional safeguards. By comparing and synthesizing approaches across different protocols and historical research, we aim to provide actionable guidelines for designing safer and more trustworthy agentic webs.

Our contributions in this work are fourfold. (1) We propose a unifying framework that delineates these six trust models in the context of multi-agent systems, highlighting their theoretical foundations in computational trust and distributed security (2) We review the state-of-the-art agent interaction protocols (A2A, AP2, ERC-8004, NANDA, etc.) and analyze how each protocol leverages one or more of these trust mechanisms in practice. (3) We critically examine how each trust model addresses (or fails to address) LLM-specific failure modes and risks, such as prompt injection attacks, susceptibility to manipulation or “sycophancy,” hallucinated outputs, deceptive strategies, emergent power-seeking behavior, and objective misalignment. We argue that certain LLM fragilities fundamentally limit the effectiveness of purely reputational or self-claimed trust approaches, necessitating a hybrid approach. (4) We outline a forward-looking research agenda and design implications, including the need for trust tiering (calibrating trust mechanisms to the risk level of tasks), combining multiple trust signals for robustness, incorporating human oversight and auditability, and addressing open questions around governance, standardization, and ethical considerations in agentic ecosystems.

\section{Background}

\subsection{Trust in Multi-Agent Systems}

Trust is a multifaceted concept that has been extensively studied in philosophy, sociology, and computer science. Philosophically, trust is often defined as a \emph{directional} and \emph{context-specific} relationship: agent A may trust agent B for task X but not necessarily for task Y \cite{Manzini2024Should}. Trust involves a belief or expectation by the trustor (the one who trusts) that the trustee (the one being trusted) has both the competence and the willingness to perform the entrusted task. Crucially, trust carries an element of \emph{vulnerability}: the trustor risks being let down or betrayed if the trustee does not fulfill expectations \cite{ONeill2002Autonomy}. This vulnerability differentiates trust from mere \emph{reliability} on predictable behavior \cite{Freiman2023Making}. For example, one can rely on a simple machine or deterministic software, but trusting an agent implies an assumption about its motives or integrity, not just its regularity of output. A trustworthy agent, then, is one that is deserving of trust–it consistently proves competent and benevolent in the relevant domain \cite{o2015trust}. In an agent society, the core challenge is “how to trust the trustworthy but not the untrustworthy” \cite{ONeill2018Linking}.

In computational terms, formalizing trust has been an ongoing endeavor since at least the 1990s. \citet{marsh1994formalising} introduced one of the first frameworks for computational trust, representing trust as a quantitative mental state that an agent can update based on experience and context. Since then, numerous models have arisen to help agents decide whom to trust in open systems \cite{Braga2019Survey}. These include reputation systems that allow agents to share impressions of each other (e.g., the ReGreT system \cite{Sabater2004EVALUATING}), probabilistic trust models that use Bayesian updating (such as the Beta reputation system \cite{josang2002beta} or TRAVOS, which computes confidence intervals for partner reliability \cite{Teacy2006TRAVOS}), and trust networks inspired by PageRank (like EigenTrust, which aggregates local trust scores in a global iterative algorithm \cite{Kamvar2003Eigentrust}). Notably, many designs aim to be collusion-resistant and Sybil-resistant. Other systems introduce whitewashing protections (preventing agents from discarding a bad reputation by rejoining under a new identity) by linking trust to persistent identities or charging entry costs. Despite these advances, fully solving trust in open agent networks remains difficult---as \citet{Friedman2001Social} noted, even an ideal reputation system only works if honest feedback is plentiful and identities cannot be cheaply faked, conditions that adversaries often undermine.

\subsection{LLM-Specific Fragilities and Trust Considerations}
While autonomous agents can in principle be built from many types of AI, the recent surge of interest in “agentic AI” has been driven by large language models. However, they also introduce unique failure modes and fragilities that complicate trust. We briefly outline these issues, as they will be referenced when evaluating trust mechanisms:

\paragraph{Prompt Injection} LLMs follow the instructions in their input prompt faithfully, which means a maliciously crafted input can inject directives that subvert the agent’s intended policy \cite{Liu2024Prompt}. This means an LLM agent might be tricked by another agent or a user into behaving badly, so any trust model that assumes an agent will strictly follow its original training or rules is vulnerable.

\paragraph{Hypersensitivity to Nudging / Sycophancy}

LLMs fine-tuned with human feedback tend to exhibit sycophantic behavior---they adapt their answers to what they think the user (or another agent) wants to hear, even if it’s not true \cite{Sharma2025Understanding}. This “nudge-susceptibility” means an agent can be influenced subtly over a conversation to adopt goals or beliefs that diverge from its initial objectives \cite{Cherep2025LLM}. In multi-agent settings, a clever adversarial agent might socially engineer a gullible LLM agent into gradually altering its behavior. This fragility undermines trust models like reputation: an agent with a good reputation could be manipulated at run-time to act against its character.

\paragraph{Hallucination}
LLMs are notorious for generating factual hallucinations, i.e. outputs that are fluent and confident-sounding but entirely incorrect or fabricated \cite{Xu2025Hallucination}. Hallucinations erode the effectiveness of self-proclaimed Claims (an agent might claim capability it doesn’t actually have) and can even fool reputation systems if others cannot verify ground truth easily. Combating hallucinations often requires external verification (e.g. cross-checking facts) which intersects with the Proof trust model.

\paragraph{Deception}
Beyond inadvertent falsehoods, sufficiently advanced agents may learn to deceive others \cite{Hubinger2024Sleeper}. Deceptive behavior directly attacks trust—especially Reputation (an agent might behave well under observation to gain reputation, then betray) and undermines naive reliance on agent self-descriptions or proof unless the proofs are comprehensive. It suggests the need for mechanisms like Stake  or tamper-proof Transparency logs to detect dishonesty.

\paragraph{Emergent Power-Seeking and Misalignment}
The most ominous concern from AI safety research is that a misaligned agent might develop instrumental goals like acquiring power or resources, and do so covertly \cite{Carlsmith2024PowerSeeking}. A misaligned LLM agent (particularly if given long-term planning ability and self-improvement loops) might identify ways to increase its influence or avoid being shutdown, actions that can be catastrophic \cite{Lynch2025Agentic}. Moreover, misalignment in goals implies that trust should not be assumed to \emph{monotonically} increase over time—an agent that was aligned yesterday might shift objectives tomorrow, so trust mechanisms must be able to reset or revoke trust quickly.

In summary, these LLM fragilities introduce trust challenges that demand a blend of security, economic, and social solutions—considerations we keep in mind while examining each trust model's strengths and weaknesses.

\section{Trust Models in Inter-Agent Protocol Design}

\begin{table*}[!htbp]
\centering
\small
\setlength{\tabcolsep}{4pt}
\renewcommand{\arraystretch}{1.2}
\newcolumntype{Y}{>{\raggedright\arraybackslash}X}
\begin{tabularx}{\textwidth}{l Y Y Y Y Y}
\toprule
\textbf{Trust Model} & \textbf{Basis of Trust} & \textbf{Strengths} & \textbf{Weaknesses} & \textbf{Mitigates LLM Issues} & \textbf{Notable Uses} \\
\midrule
\textbf{Brief} &
Third-party or self-issued credentials (verifiable). &
Quick bootstrapping of identity and roles; portable trust across contexts; cryptographically verifiable endorsements. &
Depends on issuers/authorities; requires robust revocation; static, may not reflect real-time behavior. &
Prevents simple impersonation or lying about identity/capability; does not stop runtime attacks beyond credential scope. &
NANDA AgentFacts and VCs; SSL/TLS certificates; W3C Verifiable Credentials. \\
\addlinespace
\textbf{Claim} &
Agent's self-proclaimed descriptions (AgentCard, profile). &
Lightweight, no infrastructure needed; essential for discovery and initial interfacing. &
Unverified; prone to false claims; weak incentives for truthfulness unless combined with other mechanisms; vulnerable to prompt tampering. &
Barely addresses LLM fragility—an agent can claim safety but still be misled or err internally. &
A2A AgentCards; basic API descriptions; peer agent protocols without a trust layer. \\
\addlinespace
\textbf{Proof} &
Cryptographic proofs of actions or state (signatures, zero-knowledge proofs, TEE attestations). &
High-assurance, trust-minimized verification; can preempt or catch incorrect results; enables trust in hostile settings. &
Computationally expensive; requires verifiable task specifications; TEEs have attack surfaces; limited availability. &
Strongly addresses correctness (hallucination) and some deception (proof reveals lies); does not directly solve goal misalignment. &
ERC-8004 validation registry (staked re-execution, zkML); blockchain smart contracts; TEE-based agent enclaves. \\
\addlinespace
\textbf{Stake} &
Economic collateral at risk (slashing conditions, insurance). &
Aligns incentives with honest behavior; deters bad actors via financial risk; enables automated penalties and rewards. &
Requires robust detection of misbehavior; Sybil risk if identity is cheap; may favor wealthy agents; can be gamed if stakes are mis-set. &
Discourages deceit and recklessness (agents “think twice” if large stake may be lost); does not prevent first-time or undetected misbehavior. &
ERC-8004 crypto-economic validation (stakers re-run tasks); token-curated registries; prediction markets for agent performance. \\
\addlinespace
\textbf{Reputation} &
Community feedback and history (ratings, trust scores). &
Adaptive, information-rich evaluation over time; fosters earned trust and continuous improvement; social accountability. &
Slow to build or change; susceptible to collusion, Sybil, and false reporting; cold-start problem for new agents. &
Over time filters out consistently poor or malicious agents; indirect mitigation of errors but not immediate prevention. &
EigenTrust (P2P); e-commerce seller ratings; ERC-8004 reputation registry; peer-review style networks. \\
\addlinespace
\textbf{Constraint} &
Technical limits on agent actions (sandboxing, least privilege). &
Strong safety net—contains damage regardless of agent intent; reduces need to trust agent goodwill; enforces policies strictly. &
May reduce functionality and efficiency; requires secure sandbox tech; not foolproof (sandbox escapes); a substitute for trust rather than a measure of it. &
Blocks many LLM attack vectors (cannot execute disallowed actions); ensures misaligned agents cannot exceed bounded capabilities. &
A2A security/sandbox recommendations; dockerized tool execution; OS-level app containment; tiered access control in AP2. \\
\bottomrule
\end{tabularx}
\caption{Comparison of trust models for inter-agent protocol design across basis of trust, strengths/weaknesses, LLM-related mitigations, and representative uses.}
\label{tab:trust-models}
\end{table*}

Effective trust between agents can be established through different mechanisms. We classify six major trust models and compare their characteristics in the context of agentic web protocols. Table \ref{tab:trust-models} provides a high-level comparison. Below, each model is discussed in turn, with examples and analysis of how it addresses security and LLM-specific issues.

\subsection{Brief: Endorsements and Credentials}

The Brief model grounds trust in attestations issued by trusted authorities or chains-of-trust. An agent presents signed credentials that assert properties such as identity, capability, or compliance, which others verify against issuer public keys or registries. Longstanding infrastructures—public-key certificates and web-of-trust schemes—illustrate the basic idea, and contemporary agent frameworks generalize it with verifiable credentials that bind claims to cryptographic identities and expiry policies.

This model assumes the existence of at least one trustworthy issuer and a secure binding between the credential and the agent’s cryptographic identity. It also presumes that credential semantics are sufficiently stable for the relevant decision window: credentials are issued, refreshed, and revoked on timescales that may lag behind real-time behavior. The main attack surfaces flow from these assumptions. If issuers are compromised, negligent, or captured, credentials can be misleading. If revocation is slow or fragile, credentials can outlive the agent’s good behavior. If credentials are not strongly bound to the agent’s key material, impersonation becomes possible. Finally, categorical credentials may fail to encode context, leading to overreach (for example, certification in one domain misapplied to another).

Despite these risks, Briefs excel at bootstrapping: they enable rapid, portable trust for discovery and initial contact. They mitigate LLM misrepresentation by replacing self-assertion with signed attestation; a prompt cannot conjure a valid third-party credential. They also facilitate accountability, because credentials and their verification events can be logged. The principal drawbacks are authority dependence, potential centralization, and coarse granularity: over-reliance on a small set of issuers creates choke points; credential states can lag behind behavior; and binary badges rarely reflect nuanced, dynamic competence. In practice, robust revocation, short credential lifetimes, issuer diversification, and strong identity binding are prerequisites for deploying the Brief model safely at scale.

\subsection{Claim: Self-Descriptions of Identity, Policy, and Capability}

The Claim model begins with what agents say about themselves. Protocols typically require an agent profile or “card” enumerating identity, version, skills, supported APIs, and operating policies. Such claims are necessary for discovery and routing; they advertise dynamic attributes that no external authority can continuously certify (for example, current availability or resource capacity).

Claim-based trust assumes a baseline of good faith or, at least, that misrepresentations will be uncovered downstream and discounted over time. On its own, this model is brittle in adversarial environments. Attackers can overclaim capabilities, craft deceptive profiles, or exploit ambiguous schemas. LLM agents may hallucinate competence or present inconsistent self-descriptions under prompt pressure. Even when claims are signed and fetched over secure channels to protect integrity in transit, their truthfulness remains unverified.

Where Claim shines is lightweight scalability and timeliness. It imposes the lowest infrastructural burden and supports rapidly changing or inherently subjective qualities. It also improves human and agent interpretability of intent: explicit operating policies and limitations in a profile can inform safe composition. However, Claim provides negligible direct mitigation for LLM fragilities. A model can profess safety yet be subverted by injection; it can intend to follow rules yet fail under distribution shift. Consequently, Claim should be treated as input to stronger mechanisms—filtering candidates for further vetting by credentials, proofs, reputation, or small-stake trials—rather than as a sufficient basis for critical decisions.

\subsection{Proof: Cryptographic and Verifiable Evidence}

The Proof model replaces promises and endorsements with verifiable evidence that an agent took particular actions or satisfied specified properties. Mechanisms include digital signatures and tamper-evident logs; attestations from trusted execution environments; and zero-knowledge proofs that establish correctness or compliance without revealing sensitive details. In blockchain-adjacent designs, agents may anchor hashes of actions on chain or submit zk-proofs that validate computations.

Proof-based trust presupposes verifiability: tasks must be amenable to specification and checking, or at least to attested execution in an enclave. It also relies on the soundness of the underlying cryptography and hardware. The core strength is trust minimization. Interacting parties need not know the agent’s history or reputation; they can accept or reject a transaction purely on the basis of a valid proof. For LLM agents, proofs counter several failure modes: tamper-evident logs deter denial and enable post-hoc accountability; proof-of-process or attested tool calls constrain the gap between “what the model claims it did” and “what actually executed”; and zk-proofs can certify compliance with policies (for example, “no personal data left the enclave”) without exposing internals.

Nevertheless, proofs guarantee integrity, not alignment. An agent can correctly prove that it executed a harmful policy if the policy itself is flawed. Verification scope is therefore pivotal: protocols must specify what must be proved and at what granularity. Proof systems also incur costs—circuit design, proof generation time, hardware requirements for TEEs—and introduce new single points of failure in the cryptographic stack. Side-channel and supply-chain risks in hardware attestations, denial-of-service via expensive verifications, and partial logging (where only favorable actions are proved) represent additional attack surfaces.

In practice, Proof is best deployed surgically on high-impact steps: signing outputs; attesting privileged tool invocations; proving constraint satisfaction; and anchoring audit trails. Used this way, Proof significantly elevates security against LLM deception and hallucination while keeping overhead tractable. Its limitations argue for coupling with governance elements that assert what should be proved and Constraint mechanisms that restrict what needs proving in the first place.

\subsection{Stake: Collateral, Slashing, and Incentive Alignment}

The Stake model engineers trust through skin in the game. Agents post collateral—monetary or otherwise—subject to loss if they violate protocol rules or fail to deliver contracted outcomes. Slashing conditions can be adjudicated algorithmically, by on-chain verifiers, or via human or multi-agent arbitration. Stake is attractive in open environments where identities are cheap: it imposes a cost on misbehavior and a visible signal of commitment.

Stake presumes utility-maximizing agents who care about losing collateral, reliable fault determination, and appropriately calibrated stake-to-risk ratios. It performs poorly against purely malicious adversaries willing to burn stake for damage or where potential gains from cheating exceed the maximum slash. Attack surfaces include Sybil splitting (spreading risk across many small identities), collusion in adjudication, and last-mile betrayal (accumulating reputation under small transactions and defecting on a large one). Determining what to slash for is delicate: penalizing only intentional malice leaves negligent harm unpriced; penalizing accidents risks chilling honest participation.

Stake shines when combined with verifiability and feedback. Proofs and audits provide objective grounds for slashing; reputation raises the opportunity cost of misbehavior; and claim/credential gates can modulate required stake. For LLM agents, stake introduces an economic learning signal: repeated penalties for errors or unsafe actions create incentives to adopt safer prompts, tool-use, and self-checks. However, stake is largely ex post; it cannot prevent a single catastrophic action if detection and adjudication occur later. Moreover, high stake requirements skew participation toward resource-rich actors, potentially centralizing an agent economy and erecting barriers to entry.

Designers should prefer progressive staking—small bonds for low-impact tasks, rising with privilege and potential externality—paired with transparent, appealable adjudication and anti-Sybil measures (for example, binding identities to keys with history, entry deposits, or credential prerequisites). When thoughtfully calibrated, Stake is a powerful complement that turns abstract norms into enforceable incentives.

\subsection{Reputation: Distributed Feedback and Social Signals}

The Reputation model aggregates interaction outcomes into a standing that others can query when selecting partners. Signals may be quantitative (scores, star ratings) or qualitative (reviews, endorsements), global or task-specific, and centrally stored or distributed. Reputation embraces the intuition that past behavior predicts future behavior and leverages the crowd to surface reliability and quality.

Its assumptions are well known: participants provide honest feedback; identities persist long enough for history to matter; and the environment features repeated games in which agents value future opportunities. In adversarial settings, classic vulnerabilities arise: Sybil attacks and ballot stuffing, collusion, defamation of competitors, whitewashing by discarding identities, and cold-start inequities for newcomers. Weighting schemes, reputation decay, and trust-in-the-reviewer algorithms mitigate but cannot eliminate these risks.

Reputation’s distinctive strengths are adaptivity and expressiveness. It can track multidimensional qualities, emphasize recency to reflect drift, and cover domains where formal verification is impractical or subjective judgments dominate. For LLM agents, reputation can proxy for robustness: agents that repeatedly succumb to prompt injection or hallucinate will accumulate negative feedback and be filtered out of high-value interactions. Reputation also functions as an implicit stake: agents who value their standing will refrain from opportunistic misbehavior.

However, reputation is a lagging indicator and can amplify inequalities. It neither prevents first-time catastrophes nor guarantees that a high-reputation agent will behave well when incentives flip. Over-reliance invites “reputation milking,” where an actor behaves well to amass trust and defects when the stakes are highest. Privacy and fairness concerns also arise if histories are globally visible and indelible. Consequently, reputation should be scoped (task-specific where possible), tempered (with decay and confidence intervals), and combined with controls that cap the damage any single interaction can cause.

\subsection{Constraint: Sandboxing and Capability Bounding}

The Constraint model limits what agents can do, minimizing the need to predict what they will do. By enforcing least privilege, isolating execution, mediating tool access through audited gateways, and rate-limiting sensitive operations, protocols can bound harm even when agents misbehave or fail. Constraint shifts trust from the agent to the framework.

This model assumes we can identify dangerous resources and reliably restrict them. It depends on the soundness of sandboxing technologies and policy engines, and it introduces engineering overhead and potential performance costs. The attack surfaces are familiar from systems security: sandbox escapes, confused deputy abuses of permitted interfaces, covert channels, and policy drift as capabilities evolve. A second-order risk is complacency: assuming constraints are perfect and neglecting monitoring, proofs, or incentives.

Constraint is highly effective against LLM-specific vulnerabilities. It limits blast radius for prompt injection by narrowing action surfaces and validating inputs and outputs; it precludes privilege escalation by denying access to broad system interfaces; and it enables staged onboarding, where agents graduate from read-only sandboxes to broader permissions after demonstrating good behavior. Logging and mediated I/O support auditability and human-in-the-loop overrides for high-impact actions. The principal drawback is capability throttling: over-constrained agents underperform, and discovering the right balance between safety and autonomy is non-trivial. Moreover, constraints cannot neutralize harmful content within allowed channels unless paired with content-level policies and checks.

In practice, Constraint should be dynamic and graduated: protocols can encode “trust tiers” in which higher privileges require stronger credentials, more extensive proofs, higher stake, and better reputation. This aligns operational safety with demonstrated trustworthiness.

\section{How Current Protocols Implement These Trust Models}

Having defined the trust mechanisms, we turn to concrete protocols – A2A, AP2, ERC-8004, etc. – to see how they incorporate one or more of these models. Each protocol was created with slightly different goals and assumptions, which is reflected in their trust design.
\paragraph{Google’s A2A (Agent-to-Agent) Protocol}
A2A is an open specification for inter-agent interoperability that standardizes how agents describe themselves and communicate \cite{a2a}. Concretely, each agent exposes an AgentCard—typically a JSON document at a well-known endpoint—that advertises identity, capabilities, and contact details, and it exchanges requests and responses over a secured JSON-RPC channel layered on HTTPS with mutual authentication. In trust-model terms, A2A natively privileges \emph{Claim} (the self-described AgentCard) and \emph{Constraint} (enterprise controls and least-privilege network policy), with \emph{Brief} appearing through transport-level credentials such as TLS certificates and OAuth tokens. It does not prescribe ecosystem-wide \emph{Reputation}, \emph{Stake}, or cryptographic \emph{Proof} of correct computation, and it leaves discovery, vetting, and risk policy to deployments. This design makes A2A easy to adopt in organizational settings where participants are known a priori: enterprises can gate traffic with allow-lists, bind AgentCards to domains and service accounts, route calls through API gateways, and record exhaustive logs for audit. Strengths therefore include pragmatic interoperability, compatibility with existing identity and access-management stacks, and clear operational observability. However, these same choices limit A2A in open, adversarial environments: unverified AgentCards place weight on declarations rather than evidence; absence of protocol-level staking, validation, or global feedback weakens Sybil and collusion resistance; and reliance on perimeter controls does little to mitigate LLM-specific failures such as prompt injection or sycophancy at runtime. In practice, A2A benefits from pairing with higher-assurance layers—credential briefs for identity attestation, reputation directories for discovery, proof-or-stake validation for high-impact actions, and hardened sandboxes for tool use—so that its transport and schema can operate within a richer, defense-in-depth trust fabric.

\paragraph{Agent Payments Protocol (AP2)}
AP2 is a payments-oriented standard that enables AI agents to initiate and complete commerce on behalf of users while preserving accountability \cite{ap2}. Its core abstraction is the \emph{Mandate}, a verifiable credential that captures explicit authorization and context (for example, an intent cap, a cart summary, and whether a human was present), co-signed by relevant parties and presented with each transaction. Trust-model integration is therefore explicit: \emph{Brief} and \emph{Proof} are used to bind agent actions to real entities via signed mandates and verifiable identities; \emph{Constraint} is enforced through role separation and tokenization (agents do not directly handle sensitive payment credentials and interact via scoped intermediaries); and allow-listed participants provide an initial, curated \emph{Claim/Brief} layer while the ecosystem matures toward broader domain and mTLS-based assurance. Although AP2 does not standardize \emph{Reputation} or \emph{Stake} on the wire, it anticipates their use off-path: transaction outcomes and chargebacks feed risk engines that effectively act as reputation signals, and liability models, insurance, or performance bonds can supply economic skin-in-the-game for higher-risk cohorts. The approach’s principal strengths are strong ex-ante consent capture, cryptographically auditable traces, and a clear chain of responsibility that supports dispute resolution and regulatory compliance; in short, it operationalizes “trust but verify” for financial actions. Its weaknesses arise from the same controls: early reliance on curated allow-lists creates onboarding friction and may concentrate platform power; verifiable-credential issuance and policy evaluation add latency and implementation cost; mandate proofs attest authorization, not correctness of upstream agent reasoning (so LLM hallucinations must be contained by separate constraints); and the absence of standardized staking or validator quorums means runtime misbehavior is addressed mainly via ex-post risk management rather than protocol-mandated slashing. AP2 works best as the transaction spine in a layered architecture where reputational routing and optional stake-backed validations are used to gate expensive or high-impact payments, and where agents operate under strict capability bounds to blunt prompt-level exploits.

\paragraph{Ethereum ERC-8004 (Trustless Agents)}
ERC-8004 proposes a decentralized trust layer for agent discovery and assurance built around three composable registries \cite{erc8004}. An on-chain \emph{Identity Registry} assigns each agent a persistent handle (typically an NFT) that links to off-chain metadata such as an AgentCard, enabling portable identification across contexts; a \emph{Reputation Registry} aggregates structured feedback about agent performance; and a \emph{Validation Registry} coordinates third-party checks—re-execution, TEE attestation, or zero-knowledge proofs—often collateralized by stake so that faulty or dishonest behavior can be penalized. The standard makes trust modalities explicit: agents declare “supportedTrust” capabilities (e.g., reputation, crypto-economic validation, TEE attestation), thereby advertising which evidence they can furnish. In trust-model terms, ERC-8004 integrates \emph{Claim/Brief} (self-descriptions anchored in on-chain identity and, optionally, external credentials), \emph{Reputation} (shared, queryable feedback), and \emph{Proof}+\emph{Stake} (verifiable validation with economic consequences). \emph{Constraint} is indirect but meaningful: when consequential actions are mediated by smart contracts, the agent’s latitude is bounded by program logic and policy. The architecture’s strengths are transparency, composability, and trust portability: identities and attestations are addressable; validations can be tailored per task; and reputational data can be consumed by diverse clients or even proven in privacy-preserving ways. By attaching stake to validation, the design aligns incentives for honest behavior and encourages ecosystem watchdogs. Its weaknesses reflect blockchain trade-offs and open-system realities: on-chain interactions add cost and latency; public data raises privacy concerns unless accompanied by selective-disclosure mechanisms; validators and feedback channels themselves become targets for sybil and collusive manipulation; and, critically, the benefits are optional unless counterparties \emph{require} validation or minimum reputation thresholds. Moreover, proofs attest properties of computations, not normative desirability, so misaligned but formally correct behavior remains a governance problem. ERC-8004 is most effective when embedded as a normative requirement for high-impact workflows, while lower-stakes interactions rely on identity and reputation alone.

\section{Discussion}

\subsection{Protocol Design implications}
Drawing on our comparative analysis, we derive the following design implications.

\paragraph{Tiered trust and “trustless-by-default” for high-impact actions.} Not all interactions justify heavy machinery. Systems should implement adaptive tiers (T0–T3) in which stricter controls and stronger evidence are automatically required as potential harm rises. Low-stakes activities (read-only queries, small reversible writes) can rely on Claims and Briefs to maximize throughput, while high-stakes actions escalate to Proofs (signatures, attestations, zero-knowledge where feasible), multi-party validation, and substantial Stake/insurance. Crucially, transitions between tiers must be enforced by policy: when an agent attempts an action whose blast radius exceeds its current tier, the platform intercepts and upgrades the required checks before execution.

\paragraph{Identity and Briefs as the substrate for discovery and accountability.} Verifiable identity is not the same as trustworthiness, but it is a prerequisite for traceability, audit, and remediation. Agents should expose durable identifiers and support verifiable credentials (“Briefs”) that attest to properties such as domain expertise, safety audits, regulatory licenses, or organizational affiliation. Pseudonyms may be tolerated at low tiers; high tiers should bind agents to legal entities or qualified pseudonyms to enable recourse. This encourages healthy credential ecosystems (auditors, certifiers, registries) and lets policymakers tie thresholds (e.g., transaction size) to identity requirements.

\paragraph{Hybrid by default, configurable per task.} Because verification affordances vary across domains, systems should compose multiple models—Reputation for discovery, Briefs for eligibility, Stake for incentives, Proofs and Constraint for execution guarantees—and allow policy-per-action configuration. Architecturally this implies modular “trust hooks” (proof verification, reputation lookup, staking and slashing, sandbox provisioning) and declarative policies that choose which hooks to invoke for each operation. A practical maxim follows: start trustless, then relax—assume nothing, then introduce the minimum additional assumption needed to complete the task.

\paragraph{Reputation as a layered signal, never a single gate.} Reputation is invaluable for routing and prioritization but is vulnerable to collusion, Sybil attacks, and domain shift. Use it to influence ranking, rate limits, and sampling of secondary checks—not to waive core safety guarantees. Design for multi-dimensional reputation (accuracy, responsiveness, security compliance), decay (to reflect inactivity and model updates), transitive weighting (trust of raters), and anomaly detection (to flag correlated or suspicious feedback). Couple reputation drops with automatic escalation (e.g., more frequent proofs or audits).

\paragraph{Incentive alignment via Stake and insurance.} Economic skin-in-the-game should scale with risk. Require agents (or their principals) to post bonds proportional to potential harm; slash objectively when violations occur; and explore insurance pools that underwrite agent performance. Codify slashing conditions ex ante and make them verifiable (e.g., via attested logs or challenge-response protocols). Route slashed funds to affected users or validators to incentivize oversight. Guard against gaming (e.g., orchestrated slashing) by grounding penalties in objective, reproducible fault conditions.

\paragraph{Hard Constraints and least privilege as non-negotiables for LLM agents.} Treat agent inputs as adversarial and outputs as untrusted until checked. Run effectful actions in sandboxes with ephemeral credentials, narrow scopes, and time-bounded permissions; enforce rate limits and circuit breakers; and maintain a separation between planning (LLM) and acting (tool runner with policy guardrails). Monitor for policy violations at runtime and automatically quarantine misbehaving agents or sessions. Constraints are the last line of defense when other trust layers fail.

\paragraph{Contextual, domain-specific trust zones.} Sectors differ in harm models and legal obligations. Support trust zones with domain-tailored requirements (e.g., healthcare agents must hold clinical credentials, operate under human oversight, and log to compliant archives; creative or gaming agents can tolerate lighter regimes). Provide gateways for controlled inter-zone interaction, and ensure hooks for regulatory audit and legal evidence across zones.

\paragraph{Continuous monitoring, auditability, and non-accumulating trust.} Trust should be earned repeatedly. Maintain append-only action logs and signed receipts; sample outputs for random audits; and re-baseline trust when material conditions change (new model weights, ownership changes, or security incidents). Decay stale credentials, require periodic restaking, and treat major updates as probationary periods with elevated scrutiny.

\subsection{Design guidelines: a tiered blueprint (T0–T3)}
This tiered blueprint (T0–T3) offers a risk-calibrated, modular framework for systematically applying hybrid trust models across tasks and risk levels in the agentic web.

\paragraph{T0 — Low-stakes discovery and read-only use.} Enable frictionless interoperability for negligible-impact tasks (e.g., public queries, draft generation). Rely on Claims and available Briefs for discovery; enforce soft constraints (read-only credentials, rate limits, allow-lists). Logging is best-effort; reputation may inform routing but must not gate access.

\paragraph{T1 — Moderate stakes with accountability.} Permit limited writes or small, reversible payments with explicit attribution. Require authenticated, signed intents; narrowly scoped, reversible permissions; durable receipts in secure logs. Use small, refundable bonds and minimal reputation thresholds to lift probation caps; throttle or trigger secondary checks on anomalies.

\paragraph{T2 — High stakes with strong assurance.} For materially consequential actions, adopt a “verify relentlessly” posture. Re-justify each transaction via TEE attestations, zero-knowledge or interactive proofs, or quorum validation; enforce deny-by-default, fine-grained, time-boxed privileges with continuous monitoring. Calibrate stake/insurance to worst-case loss; maintain immutable audit trails; reserve human review for exception paths. Reputation may rank eligible agents but never substitutes for proofs.

\paragraph{T3 — Critical or life-critical with multi-layer oversight.} In safety-critical or ethically sensitive domains, stack all mechanisms: regulatory-grade credentials and institutional accountability; redundant agents or multi-signature approvals; human-in/on-the-loop gating with physical/procedural fail-safes; non-overridable hard limits; comprehensive, privacy-preserving observability. Extend liability beyond agent stakes to organizational insurance; engineer for graceful degradation and rapid intervention.

Modularity and invariants. Tiers are composable rather than siloed: moving upward adds evidence, incentives, and containment rather than merely “more of the same.” Two invariants span all tiers: (i) least privilege at the capability boundary (minimal, time-boxed powers), and (ii) evidence-first accountability at the audit boundary (signed state and reproducible logs enabling independent verification). Adopting this blueprint as a default, with local refinements, preserves resilience and accountability under adversarial conditions.
\section{Conclusion}

The near future of the agentic web will be shaped by protocols that treat hybrid trust as infrastructure: beginning with verification and containment, layering aligned incentives and institutional accountability, and only then exploiting social signal to regain efficiency. A tiered, composable, and continuously recalibrated approach enables low-friction exploration where impact is negligible and “verify relentlessly” where stakes are high, allowing autonomous agents to transact with the assurance we expect of conscientious human institutions.

\bibliography{reference}

\appendix

\section{Disclosure of the usage of LLM}
We used ChatGPT (GPT5 model \cite{openai2024gpt}) to facilitate the writing of this manuscript. The usage includes:
\begin{itemize}
    \item Turn Excel format tables into LaTeX format tables
    \item Correct grammar mistakes and spelling
    \item Deep research for various protocols
    \item Polish the existing writing 
\end{itemize}

\end{document}